\documentstyle[twoside,fleqn,espcrc2]{article}
\begin{document}
\vskip -3cm
\hskip 11.5cm
\vbox{\hbox{POLFIS-TH 10/95}
\hbox{CERN-TH/95-244}\hbox{hep-th/9509160}
}
\begin{center}
{\LARGE The Symplectic Structure of N=2 Supergravity \\
\vskip 1.5mm
 and its Central Extension
  $^*$ }\\
\vfill
{\large  A. Ceresole$^1$,
R. D'Auria$^1$ and S. Ferrara$^3$
  } \\

\vfill
{
$^1$ Dipartimento di Fisica, Politecnico di Torino,\\
 Corso Duca degli Abruzzi 24, I-10129 Torino\\
and Istituto Nazionale di Fisica Nucleare (INFN) - Sezione di Torino, Italy\\
\vspace{6pt}
$^2$ CERN Theoretical Division, CH 1211 Geneva, Switzerland\\
and UCLA Physics Department, Los Angeles CA, USA\\
\vspace{6pt}
}
\end{center}
\hskip 1.5cm

\vfill
\begin{center}
{\large\bf Abstract}
\end{center}
{\large
We report on the formulation of $N=2$ $D=4$ supergravity coupled to
$n_V$ abelian vector multiplets in presence of electric and magnetic charges.
General formulae for the (moduli dependent) electric and magnetic charges
for the $n_V+1$ gauge fields are given which reflect the symplectic
structure of the underlying special geometry. The specification to Type IIB
strings compactified on Calabi-Yau manifolds, with gauge group
$U(1)^{h_{21}+1}$ is given.
}
\vspace{2mm}
\begin{center}
{\large\sl Contribution to the Proceedings of the Trieste Conference on\\
 ``S Duality and Mirror Symmetry''\\
 June 1995}
\end{center}
 \vfill \hrule width 3.cm
{\footnotesize
\noindent
$^*$ Supported in part by DOE grant
DE-FGO3-91ER40662, Task C.,
by EEC Science Program SC1*CT92-0789 and INFN.}

\vfill
\eject
\newcommand{\bfone}{\relax{\rm 1\kern-.35em 1}}
\newcommand{\IC}{{\relax\,\hbox{$\inbar\kern-.3em{\mss C}$}}}
\newcommand{\IR}{\relax{\rm I\kern-.18em R}}
\newcommand{\ZZZ}{\relax\ifmmode\mathchoice
{\hbox{\cmss Z\kern-.4em Z}}{\hbox{\cmss Z\kern-.4em Z}}
{\lower.9pt\hbox{\cmsss Z\kern-.4em Z}}
{\lower1.2pt\hbox{\cmsss Z\kern-.4em Z}}\else{\cmss Z\kern-.4em
Z}\fi}
\newcommand{\La}{\Lambda}
\newcommand{\Si}{\Sigma}
\newcommand{\cF}{{\cal F}}
\newcommand{\cN}{{\cal N}}
\newcommand{\cS}{{\cal S}}
\newcommand{\cT}{{\cal T}}
\newcommand{\cR}{{\cal R}}
\newcommand{\cL}{{\cal L}}
\newcommand{\cM}{{\cal M}}
\newcommand{\cG}{{\cal G}}
\newcommand{\cD}{{\cal D}}
\newcommand{\cI}{{\cal I}}
\newcommand{\cZ}{{\cal Z}}
\newcommand{\half}{{1\over2}}
\newcommand{\del}{\partial}
\newcommand{\re}{{\rm Re\,}}
\newcommand{\im}{{\rm Im\,}}
\newcommand{\jb}{{\bar \jmath}}
\newcommand{\beq}{\begin{equation}}
\newcommand{\eeq}{\end{equation}}
\newcommand{\lqn}{\lefteqn}

\font\cmss=cmss10 \font\cmsss=cmss10 at 7pt
\newcommand{\ZZ}{\relax\ifmmode\mathchoice
{\hbox{\cmss Z\kern-.4em Z}}{\hbox{\cmss Z\kern-.4em Z}}
{\lower.9pt\hbox{\cmsss Z\kern-.4em Z}}
{\lower1.2pt\hbox{\cmsss Z\kern-.4em Z}}\else{\cmss Z\kern-.4em
Z}\fi}

\newcommand{\ttbs}{\char'134}
\newcommand{\AmS}{{\protect\the\textfont2
  A\kern-.1667em\lower.5ex\hbox{M}\kern-.125emS}}

\hyphenation{author another created financial paper re-commend-ed}
\hyphenation{Work sup-ported in part by EEC Sci-ence Pro-gram}
\hyphenation{Ge-om-e-try}

\title{The Symplectic Structure of N=2 Supergravity and its Central
       Extension\thanks{ Supported in part by DOE grant DE-FGO3-91ER40662,
                         Task C., EEC Science Program SC1*CT92-0789
                         and INFN.}
                        }

\author{A. Ceresole\address{Dipartimento di Fisica, Politecnico di Torino,\\
        Corso Duca Degli Abruzzi 24, Torino I-10129, Italy} ,
        R. D'Auria{\hbox{$^{\rm a}$}} %
        and
        S. Ferrara\address{Cern, 1211 Geneva 23 , Switzerland
        }}


\begin{abstract}
We report on the formulation of $N\!=\! 2$, $D\!=\! 4$ supergravity coupled to
$n_V$ abelian vector multiplets in presence of electric and magnetic charges.
General formulae for the (moduli dependent) electric and magnetic charges
for the $n_V+1$ gauge fields are given which reflect the symplectic
structure of the underlying special geometry.
Model independent sum rules obeyed by these charges are obtained.
 The specification to Type IIB
strings compactified on Calabi-Yau manifolds, with gauge group
$U(1)^{h_{21}+1}$, is given.
\end{abstract}

\maketitle

\section{INTRODUCTION}

Recent developments towards a deeper understanding of the exact quantum
structure of supersymmetric non abelian gauge theories
\cite{SW1,KLTY,faraggi}
and their extension to gravity couplings
\cite{CDF,CDFVP}, including   superstring theories\cite{Wdy},
have revealed that many aspects of nonperturbative physics, in the
infrared regime, can be exactly analysed in virtue of  their supersymmetric
structure.

In particular, the peculiar r\^ole played by the dilaton-axion complex
scalar field $S$ in 4D heterotic strings, allows one to extend the powerful
non-renormalization theorems of rigid renormalizable gauge theories to the
string case
\cite{AFGNT,DKLL}
. $N=2$ theories are specially interesting examples because they
exhibit (unlike $N\geq 3$ theories) non trivial dynamics, but nevertheless
they may be exactly  solved by making some hypothesis about their
nonperturbative behaviour.

Exact solutions of the infrared regime of both $N=2$ super Yang-Mills
theories
\cite{SW1,KLTY}
 and their string analogue
\cite{FHSV}
 have been the subject of intensive
investigation over the last year
\cite{dieci}
, and spectacular results
\cite{stromco,KLT,AGNT,vafa}
substantiating
 proposed assumptions about the exact solutions have been obtained.

For $N=2$ Yang-Mills theories coupled to gravity, which are supposed to
be a low energy manifestation of $N=2$ 4D heterotic strings,
 the nonperturbative physics
of the low energy effective action can be inferred by the assumption that
a given non-abelian theory, with gauge group $G$ of rank $r$, in the
particular phase where the gauge group is broken to $U(1)^r$ is equivalent,
at the nonperturbative level, to a Type IIB supergravity\cite{KLTY}
 (which is the low
energy theory for Type IIB strings) compactified on a Calabi--Yau manifold
\cite{CDF}
(or its mirror
\cite{essays}) with Hodge numbers $h_{11}= n_H-1$ , $h_{21}= n_V=r+1$. Here
$(n_V,n_H)$ denote the number of vector multiplets and hypermultiplets,
neutral with respect to the abelian gauge group $U(1)^r$ on the heterotic
side. Since $n_V$ includes the (dual to the heterotic)
dilaton-axion vector multiplet, it follows that
$n_V\geq 1$ ($h_{21}\geq 1$) and also $n_H\geq 1$, as on the Type II side
there is always the universal hypermultiplet
\cite{seib,CFG}
 which corresponds to the Type II
string dilaton.

If one would like to compare string theories with rigid theories
\cite{BCDFF,CLM,KKLMV,antop}
 with a rank
$r$ gauge group, then $n_V=r+1$ (because the dilaton-axion degree of
freedom is frozen in the rigid limit) and this means that $h_{21}=r+1\geq 2$.
Since the overall gauge group is $U(1)^{r+2}$ (including the graviphoton),
an example of
a new nonperturbative stringy phenomenon is the possibility of having
states which have in addition electric and magnetic charges with respect to
all $r+2$ vector fields, as it occurs in higher N theories, under the
assumption of S duality or more generally U duality
\cite{schwarz,huto1}.

In section 2 we will summarize some basic formulae for the underlying
moduli space of $n_V$ complex scalar fields\cite{spec1,specs}
 of vector multiplets
\cite{spec}. It is
the geometrical properties of this space (special geometry) which
determine the low energy couplings of vector fields as well as the central
extension
\cite{savoy}
 of the $N=2$ supersymmetry algebra. These quantities receive both
perturbative and non perturbative corrections
\cite{AFGNT,DKLL}
 on the heterotic side and are
supposed to be exactly computed on the Type II side
\cite{FHSV,stromco}, at the classical level,
under the hypothesis of string-string duality
\cite{Duff,Wdy}
 and the identification
\cite{FHSV,dieci}
 of the
``dual'' dilaton-axion degree of freedom with one particular element of the
$H_{21}$ cohomology. Non perturbative phenomena must also occurr on the
Type II side, since the central extension of the $N=2$ algebra implies the
existence of electric and magnetic charged states carrying Ramond-Ramond
charges
\cite{huto1}
 together with the fact that the hypermultiplet moduli space should
also undergo quantum corrections to match the classical quaternionic space
on the heterotic side
\cite{stromco,BBS}. We will not discuss the hypermultiplet dynamics
\cite{bawi} in
this report. Such corrections may be viewed as particular wrappings of 10D
p-branes
\cite{BBS,town1}
 ($p=1,3,5$) around the Calabi-Yau manifold and therefore  are
non-perturbative in nature at the string level. In section 3
we will give general formulae for the moduli dependent charges of the
$U(1)^{n_V+1}$ gauge fields and some general relations that such charges
obey as a consequence of special geometry and its symplectic structure.
Finally, the application of these results to the compactification of
Type IIB strings on Calabi-Yau threefolds is given in section 4.

\section{SPECIAL GEOMETRY AND ITS SYMPLECTIC STRUCTURE}

Let us consider here the space of the $n_V$ complex scalar fields $z^i$ of
vector multiplets coupling\cite{spec} to
$N=2$ supergravity\cite{fvanN} .
 The local structure of this space is that of a
K\"ahler-Hodge manifold with a K\"ahler metric $G_{i\jb}$ whose curvature
satisfies the ``special geometry'' constraint
\beq
R_{i\jb l \bar m}=G_{i\jb} G_{l\bar m}+G_{i \bar m} G_{l\jb}-C_{ilp}
\bar C_{\jb \bar m\bar p} G^{p\bar p}\ .
\label{duno}
\eeq
Eq. (\ref{duno}) is a consequence of the $N=2$ local supersymmetry algebra,
and can actually be derived from the symplectic structure dictated by the
Bianchi identities. Define the (covariantly holomorphic) symplectic sections
of the Hodge bundle $\cal L$
\begin{eqnarray}
\lqn{ V = (L^\La,M_\La)\ \ \ \ \ \ \ \ \ \ \La=0,\ldots,n_V } \nonumber\\
&& D_{\bar \imath} V = (\del_{\bar\imath}-\half\del_{\bar\imath} K)V=0
\label{ddue}
\end{eqnarray}
such that
\beq
i <V,\bar V>=i(\bar L^\La M_\La-\bar M_\La L^\La)=1\ .
\label{dtre}
\eeq
Defining $U_i=D_iV=(\del_i+\half \del_i K)V=(f_i^\La,h_{i\La})$,
one finds
\beq
D_i U_j=i C_{ijk} G^{k\bar k}\bar U_{\bar k}\ ,
\label{dqua}
\eeq
where $C_{ijk}$ is a covariantly holomorphic section of
 $(T^*)^3\otimes\cL^2$,
totally symmetric in its indices. It was shown in ref \cite{specs} that eqs.
(\ref{duno})-(\ref{dqua}) define a flat symplectic connection.
{}From the above  relations  it is
possible to derive  some holomorphic identities which are equivalent to
Picard-Fuchs equations whenever $V$ is associated to the periods of some
Calabi-Yau  manifold
\cite{spec1,CDFLL}
. Eq. (\ref{duno}) can be solved by setting
\cite{CDFVP}
\begin{eqnarray}
M_\La=\cN_{\La\Si} L^\Si\ \ ,\ \  h_{i\La}=\bar\cN_{\La\Si} f^\Si_i\ ,
\nonumber\\
\im \cN_{\La\Si} L^\La \bar L^\Si=-\half ,
\label{dcin}
\end{eqnarray}
where $\cN_{\La\Si}$ is a complex symmetric matrix. Then one has
\beq
<V,U_i>=<V,U_{\bar \imath}>=0\ ,
\label{dsei}
\eeq
\beq
G_{i\jb}=-i <U_i, U_\jb>\ ,
\label{dset}
\eeq
\beq
C_{ijk}=<D_i U_j, U_k>\ .
\label{dott}
\eeq
Moreover, $\cN_{\La\Si}$ can be obtained in terms of the two
$(n_V+1)\times (n_V+1)$ matrices
\beq
f_I^\La=(f_i^\La,\bar L^\La)\ \ ,\ \ h_{I\La}=(h_{i\La},\bar M_\La)
\label{dnov}
\eeq
from the relation
\beq
\bar \cN_{\La\Si}=h_{I\La}(f^{-1})^I_\La\ .
\label{ddie}
\eeq
Another useful relation is
\beq
U^{\La\Si}=f_i^\La G^{i\jb} f_\jb^\Si=
-\half (\im \cN)^{\La\Si}-\bar L^\La L^\Si\ .
\label{dund}
\eeq
By setting
\beq
L^\La=e^{{K(z,\bar z)}\over2} X^\La(z)\  , \ M_\La=e^{{K(z,\bar z)}\over2}
 F_\La (z)
\label{ddod}
\eeq
it follows that $(X^\La,F_\La)$ are holomorphic sections of a line bundle,
and all previous formulae can be written in terms of such sections. The
K\"ahler potential is
\beq
K=-\ln i<\Omega,\bar \Omega>\ ,
\label{dtrd}
\eeq
where $\Omega=(X^\La,F_\La)=e^{-K/2} V$. From eq. (\ref{dsei}) one finds
\beq
X^\La \del_i F_\La-\del_i X^\La F_\La=0\ .
\label{dqat}
\eeq
Note that under K\"ahler transformations $K\to K+f+\bar f$ and
$\Omega\to\Omega e^{-f}$. Since $X^\La\to X^\La e^{-f}$, this means that we
can regard, at least locally, the $X^\La$ as homogeneous coordinates on
the K\"ahler manifold\cite{spec1,specs}, provided the matrix
\beq
e^a_i(z)=\del_i(X^a/X^0)\ \ \ a=1,\ldots,n_V
\label{dqin}
\eeq
is invertible. In this case, $F_\La=F_\La(X)$ and because of eq. (\ref{dqat})
and the fact that $X^\Si\del_\Si F_\La=F_\La$, one then has
\beq
F_\La (X)=\del_\La F(X)\ \ \ ,\ \  \ X^\Si\del_\Si F=2F\ .
\label{dsed}
\eeq
$F(X(z))$ is the prepotential of $N=2$ supergravity vector multiplet
couplings \cite{spec}
and ``special coordinates'' correspond to a coordinate choice for which
\cite{spec1,specs,CDFLL}
\beq
e^a_i=\del_i(X^\La/X^0)=\delta_i^a\ .
\label{ddic}
\eeq
This means $X^0=1$, $X^i=z^i$. Note that under coordinate transformations,
the sections $\Omega$ transform as
\beq
\Omega'=e^{-f_\cS (z)}\Omega\ ,
\label{ddoc}
\eeq
where $\cS=\pmatrix{A&B\cr C & D\cr}$ is an element of $Sp(2n_V+2,\IR)$,
\begin{eqnarray}
A^TD-C^TB&=&\bfone\ ,\nonumber\\
A^TC-C^TA&=&B^TD-D^TB=0\ .
\label{ddov}
\end{eqnarray}
Since $F=\half X^\La F_\La$, this implies that
\begin{eqnarray}
\lefteqn{\tilde F(\tilde X)=F(X)+X^\La(C^TB)^\La_\Si
F_\Si\nonumber}\\
&&+\half X^\La (C^TA)_{\La\Si} X^\Si
+\half F_\La(D^TB)^{\La\Si} F_\Si
\label{dven}
\end{eqnarray}
where
\beq
\tilde X=(A+B\cF) X\ \ ,\ \ \cF=F_{\La\Si}=
{{\del^2 F}\over{\del X^\La\del X^\Si}}\ .
\label{dvun}
\eeq
We also note that
\beq
\tilde\cN(\tilde X,\tilde F)=(C+D\cN
)(A+B\cN
)^{-1}\ ,
\label{dvdu}
\eeq
a relation that simply derives from its definition, eq. (\ref{ddie}) .
It is useful to give formulae which relate the two symmetric matrices
$\cN_{\La\Si}$ , $F_{\La\Si}$ which exist whenever (\ref{dqin}) is fullfilled.
We first note that $\im \cN_{\La\Si}$, $\re \cN_{\La\Si}$ are related to the
kinetic and topological term $F^2$ and $F\tilde F$ of vector fields
respectively in the low energy $N=2$ supergravity lagrangian. Thus, a
physical requirement is that
\beq
\im \cN_{\La\Si}<0\ .
\label{dvtr}
\eeq
This is actually a consequence of the positivity of the K\"ahler metric
$G_{i\jb}$ and eq. (\ref{dset})
\beq
G_{i\jb}=-2\,\im\cN_{\La\Si} f^\La_i f^\Si_\jb\ .
\label{dvqu}
\eeq
Moreover, if $\cF$ exists, then it follows that
\beq
\cN_{\La\Si}=\bar F_{\La\Si}-2i\ \bar T_\La \bar T_\Si (L\,\im\cF\, L)\ ,
\label{dvci}
\eeq
where
\begin{eqnarray}
  T_\La =-i{{(\im\cF\bar L)_\La}\over{\bar L\ \im\cF\ \bar L}} &=& 2i (
\,\im\cN\  L)_\La\  ,\nonumber\\
   L\, \im \cF\,\bar L &=&-\half  \hfill\nonumber\\
  T_\La\bar L^\La &=& -i\,  \hfill\nonumber\\
  4\bar L\, \im \cF\,\bar L&=&(L\, \im\cN\, L)^{-1}\ .\hfill
\label{dvse}
\end{eqnarray}
{}From eqs. (\ref{dqua}), (\ref{dund}) we get
\begin{eqnarray}
G_{i\jb}&=&2\, \im F_{\La\Si}\, f_i^\La f_\jb^\Si\hfill\nonumber\\
U^{\La\Si}&=&\half (\im \cF^{-1})^{\La\Si}+L^\La \bar L^\Si\nonumber\\
&=&\cT^\La_I G^{IJ}\bar\cT^\Si_J\ ,
\label{dtru}
\end{eqnarray}
where $\cT_I^\La$, $G^{IJ}$ are $(n_V+1)\times(n_V+1)$ matrices
\begin{eqnarray}
\cT^\La_I&=&(\cT_i^\La,\cT_0^\La=L^\La)\,\nonumber \\
G^{IJ}&=&(G^{ij}=G^{i\jb},G^{i0}=0,G^{00}=-1)\ .
\label{dtrdi}
\end{eqnarray}
Eq. (\ref{dtru}) implies, because of eq. (\ref{dvtr}),
that $\im \cF$ is a matrix with $n_V$
positive and one negative eigenvalue. Note that $U^{\La\Si}$ is a rank
$n_V$ matrix since it annihilates the vector $T_\La$, $\bar T_\Si$,
\beq
T_\La U^{\La\Si}=U^{\La\Si}\bar T_\Si=0\ .
\label{dvno}
\eeq
We will see in section 3 that $T_\La$ is the graviphoton projector.
{}From eq. (\ref{dtru}) we can further compute
\beq
\left[ {\det }\, 2\,\im\cF\right]^{-1}={\det }(U^{\La\Si}-L^\La \bar
L^\Si)\ ,
\label{dtrn}
\eeq
and using (\ref{dtrdi}) we find
\beq
{\det }\, 2\, \im\cF=-{\det } G_{i\jb}\mid{\det }\cT^\La_I\mid^{-2}\ .
\label{dtrt}
\eeq
In virtue of simple properties of determinants, it can be easily seen that
\beq
\det \cT^\La_I=e^{K/2\ (n_V+1)}(\det e)(X^0)^{n_V+1}\ ,
\label{dtrq}
\eeq
with $e^a_i$ given by eq. (\ref{ddic}),
and hence
\begin{eqnarray}
\lefteqn{\mid\det\cT^\La_I\mid^2=\nonumber}\\
&&e^{K(n_V+1)}(X^0\bar X^0)^{n_V+1}
\mid \det e\mid^2\ .
\label{dtrc}
\end{eqnarray}
It then follows that\cite{spec}
\beq
\del_i\del_\jb\ln\det\im\cF=\ln\det G_{i\jb}-(n_V+1)G_{i\jb}\ .
\label{dtrs}
\eeq
Since on a K\"ahler manifold  the Ricci tensor can be written as
\beq
R_{i\jb}=\del_i\del_\jb\ln\det G_{i\jb}\ ,
\label{dtse}
\eeq
one has
\beq
\del_i\del_{\bar\jmath}\ln\det\im\cF=-C_{ilp}\bar
C_{\bar\jmath\bar l\bar p}G^{l\bar l}
G^{p\bar p}\ .
\label{dtot}
\eeq
Thus  eq. (\ref{dtrq}) becomes
\beq
\det\im 2\cF(t,\bar t)=-\det G_{a\bar b}(t,\bar t)e^{K(t,\bar t)(n_V+1)}\ ,
\label{dtno}
\eeq
where $t^a={{X^a}\over{X^0}}$, $F(X^\La)=(X^0)^2 f(t^a)$, $G_{a\bar b}=\del_a
\del_{\bar b} K(t,\bar t)$, and
\beq
e^{-K(t,\bar t)}=i\left[ 2f-2\bar f+(\bar t^a-t^a)(f_a+\bar f_a)\right]\ .
\label{dquat}
\eeq

\section{ELECTRIC-MAGNETIC DUALITY, CENTRAL EXTENSION
AND ELECTRIC-MAGNETIC CHARGE RELATIONS}

In the previous section we have described the geometric structure
of the metric of scalar fields. The metric $G_{i\jb}$ appears in the
scalar kinetic term of the effective action
\beq
-G_{i\jb}\del_\mu z^i \del_\nu z^\jb g^{\mu\nu} \sqrt{-g}\ .
\label{tuno}
\eeq
The symmetric matrix $\cN_{\La\Si}$ appears in the vector part of the action
(we set the fermions to zero)
\beq
\im \cF^{-\La}\bar\cN_{\La\Si}\cF^{-\Si}=\im\cF^{-\La} \cG^-_\La\ ,
\label{tdue}
\eeq
where $\cG^-_\La=\bar\cN_{\La\Si}\cF^{-\Si}$. The symplectic structure of
the equations of motion comes by defining the $Sp(2n_V+2)$ symplectic
(antiselfdual) vector field strength
\beq
\cZ^-=(\cF^{-\La},\cG^-_\La )
\label{ttre}
\eeq
and writing, in form language
\beq
d\,\re \cZ^-=0
\label{tqua}
\eeq
in a source free theory. In presence of electric and magnetic sources we
can write
\beq
\int_{S_2} \cF^\La =n^\La_m\ ,
\label{tcin}
\eeq
\beq
\int_{S_2} \cG_\La = n^e_\La\ ,
\label{tsei}
\eeq
where the symplectic vector $\re \cZ^-$ is
\begin{eqnarray}
\re \cZ^- &=&(\cF^\La , \cG_\La)\nonumber\\
\cG_\La &=& \re \cN_{\La\Si} \cF ^\Si +\half\,
\im \cN_{\La\Si} \tilde \cF^\Si \ .
\label{tset}
\end{eqnarray}
Since $(n^\La_m , n^e_\La )$ are integers, eqs. (\ref{tcin}), (\ref{tsei})
are only covariant under $Sp(2n_V+2;\ZZ )$ rotations. We can now define
two symplectic invariant combinations\cite{CDF,BCDFF}
 of the symplectic field strength
vector $\cZ^-$:
\beq
T^-=-<\cZ^-,V>=T_\La \cF^{-\La}\ ,
\label{tott}
\eeq
with $V$ and $T_\La$ defined by eqs (\ref{ddue}), (\ref{dvse}),
\beq
\cF^{-i}=-<\cZ^-, D_\jb \bar V> G^{i\jb}\ .
\label{tnov}
\eeq
Because of eq. (\ref{dcin}) we also have
\beq
<\cZ^-, \bar V>=<\cZ^-, D_j V>=0\ .
\label{tdie}
\eeq
It then follows that\cite{BCDFF,FKLZ}
\beq
-\half \int_{S_2} T^-= Z= L^\La n_\La^e-M_\La n^\La_m\ ,
\label{tund}
\eeq
\beq
-\half \int_{S_2} \cF^{+\jb} G_{i\jb}= D_i Z= Z_i\ .
\label{tdod}
\eeq
The explicit expression for $T^-$, $\cF^{-i}$ are
\beq
T^-=2i\,\im\cN_{\La\Si} L^\La \cF^{-\Si}\ ,
\label{ttrd}
\eeq
\beq
\cF^{-i}=2i G^{i\jb} \im \cN_{\La\Si} \bar f^\La_\jb \cF^{-\Si}\ .
\label{tqut}
\eeq
We may define the field strength $\hat\cF^{\La}$ orthogonal to the graviphoton,
\beq
\hat \cF^- T_\La=0\ \ {\rm i.e.}\ \ \hat\cF^{-\La}=\cF^{-\La}-i\,\bar L^\La
T_\Si \cF^{-\Si}
\label{tqin}
\eeq
and notice that $\cF^{-i}$ is orthogonal to the graviphoton since in
(\ref{tqut}) $\cF^{-\La}\to \hat\cF^{-\La}$ in virtue of the fact that
\beq
\im \cN_{\La\Si} \bar f^\La_\jb \, \bar L^\Si=0\ .
\label{tsed}
\eeq
The objects defined by eqs. (\ref{tott}), (\ref{tnov}) have the physical
meaning of being the (moduli-dependent) vector combinations which appear in
the gravitino and gaugino supersymmetry transformations respectively
\cite{spec}. Indeed, setting to zero for simplicity the fermion bilinears
and the $N=2$ generalization of the Fayet-Iliopoulos term, we have
\begin{eqnarray}
\delta \psi_{A\mu} &=& \cD_\mu \epsilon_A +\epsilon_{AB} T_\Lambda \cF^{-\La}
_{\mu\nu}\gamma^\nu \epsilon^B\nonumber\\
\delta \lambda^{iA} &=& i\gamma^\mu \del_\mu z^i \epsilon^A+
{i\over2} \cF^{i-}_{\mu\nu}\gamma^{\mu\nu}\epsilon_B\epsilon^{AB}\ ,
\label{ricca}
\end{eqnarray}
where $\lambda^{iA}$, $\psi_{A\mu}$ are the chiral gaugino and gravitino
fields, $\epsilon_A\ ,\ \epsilon^A$ are the chiral and antichiral
supersymmetry parameters respectively, $\epsilon^{AB}$ is the $SO(2)$
Ricci tensor.

Eq. (\ref{tund}) define the central extension of the local supersymmetry
algebra,
$Z$ being the central charge. Eq. (\ref{tdod}) defines in a geometrical way
the charges of the other field strength vectors, orthogonal to the graviphoton.
Note that the charges $(Z, Z_i)$ are in correspondence with the real charges
$(n^\La_m,n^e_\La)$, but refer to the supermultiplet eigenstates, which are
moduli dependent. The charges $(Z,Z_i)$ satisfy two model independent
relations which follow from their definition and eqs. (\ref{dund}),
(\ref{dtru})
\beq
\mid Z\mid^2+\mid Z_i\mid^2=-\half P^t \cM(\cN) P\ ,
\label{tdic}
\eeq
\beq
\mid Z_i\mid^2-\mid Z\mid^2=\half P^t \cM (\cF) P\ ,
\label{tdoc}
\eeq
\begin{eqnarray}
\lqn{P^t\cM(\cN)P =}\nonumber\\
\lqn{(n^e_\La-\bar\cN_{\La\Si} n_m^\Si) \im\cN^{-1\La\Si} (n^e_\Delta
-\cN_{\Delta\Gamma}n^\Gamma_m)}
\end{eqnarray}
Here $P=(n^\La_m,n^e_\La)$ and $\cM(\cN)$ is the real symplectic matrix
\begin{eqnarray}
\lqn{\cM(\re \cN,\im\cN)=}\nonumber \\
&&\cR^t (\re \cN)\cD (\im\cN)\cR (\re\cN) ,
\label{tven}
\end{eqnarray}
where
$\cR (\re \cN)=\pmatrix{\bfone&0\cr -\re\cN &\bfone}$,
$\cD(\im\cN)=\pmatrix{\im \cN&0\cr 0&\im\cN^{-1}}$. The same matrix appears
in (\ref{tdoc}) with $\cN\to\cF=F_{\La\Si}$. It is evident that eqs.
(\ref{tdic}),(\ref{tdoc}) reflect the fact that $\im\cN$ is negative definite
and that $\im\cF$ has $n_V$ positive and one negative eigenvalue. This
comes from the signature of the quadratic form on the right-hand-side
and further noticing that $\re\cN (\re\cF)$ don't play any role because
$\cR$ can be included in the vector $P$
\beq
P_\cR =\cR P\ ,
\label{tvuno}
\eeq
so that one gets $P^t \cM P= P^t_\cR \cD P_\cR$ which is manifestly a
quadratic form with negative signature or $(n_V,1)$ signature for the two
cases at hand. As an illustrative example, let us compute some consequences
of formulae (\ref{tdic}),(\ref{tdoc}) for the homogeneous special manifold
$SU(1,n)/SU(n)\times U(1)$ and ${{SU(1,1)}\over{U(1)}}\times{{SO(2,n)}\over
{SO(2)\times SO(n)}}$. If we take, in the first case
\beq
F(X)=-{i\over2} (X_0^2-X^2)=-{i\over2} X^\La \eta_{\La\Si} X^\Si\ ,
\label{tdnov}
\eeq
then we get
\beq
\mid Z_i\mid^2-\mid Z\mid^2=-\half (n^e_\La n^{e\La}+n_m^\La n_{m\La})\ .
\label{tvent}
\eeq
If we take now the other case, we cannot compute $F_{\La\Si}$ since it does
not exist in the $SO(2,n)$ covariant basis given by $(X^\La,F_\La=
S\eta_{\La\Si} X^\Si)$,  with $\eta={\rm diag}(++,---)$. By explicit knowledge
of
\beq
\cN_{\La\Si}=(S-\bar S)(\Phi_\La \bar\Phi_\Si+\bar\Phi_\La\Phi_\Si)+\bar
S\eta_{\La\Si}\ ,
\label{tvun}
\eeq
with $\Phi_\La=X_\La/(X^\La \bar X_\La)^{\half}$, ($X^\La X_\La=0$) we get
\beq
\im\cN_{\La\Si}={i\over2} (S-\bar S)L_{\La\Si}\ ,
\label{tvdue}
\eeq
with $L_{\La\Si}=\eta_{\La\Si}-2(\Phi_\La\bar\Phi_\Si+\bar\Phi_\La\Phi_\Si)$,
$L_{\La\Si}L^{\Si\Delta}=\delta^\Delta_\Si$ so we finally obtain
\beq
\mid Z_i\mid^2+\mid Z\mid^2=-n_\La n_\Si L^{\La\Si}{{ \mid m_1-m_2 S\mid^2}
\over{i(S-\bar S)}}
\label{tvtre}
\eeq
and we have set
\beq
n^\La_m=m_2 n^\La\ \ ,\ \ n^e_\La=m_1 n_\La\ .
\label{tvqua}
\eeq
Since we have that
\beq
\mid Z\mid^2={1\over{i(S-\bar S)}} n_\La n_\Si \Phi^\La \bar\Phi^\Si
\mid m_1-m_2 S\mid^2
\label{tvcin}
\eeq
we get finally
\beq
\mid Z_i\mid^2-3\mid Z\mid^2=-n_\Si n^\Si {{\mid m_1-m_2 S\mid^2}\over
{i(S-\bar S)}}\ .
\label{tvsei}
\eeq
The factor $3$ can be understood because $i$ runs over $n+1$ values and one
has to match
the $(n,2)$ signature of the right hand side.

\section{TYPE IIB STRINGS ON CALABI-YAU THREEFOLDS}

We can specify the previous formulae to the case of Type IIB 10D supergravity
 on Calabi-Yau manifolds. In this case $n_V=h_{21}$ and the holomorphic section
\cite{CFG,spec1,specs}
\beq
(X^\La,F_\La)
\label{quno}
\eeq
is related to the $(3,0)$ form $\Omega$
\beq
\Omega=X^\La \alpha_\La-F_\La \beta^\La\ ,
\label{qdue}
\eeq
where $(\alpha,\beta)$ is a basis in $H^3$ with $\int \alpha_\La\wedge\beta^\Si
=\delta_\La^\Si$, $\int \alpha_\La\wedge \alpha_\Si=\int\beta^
\La\wedge\beta^\Si=0$. The self-dual five form of the Type IIB supergravity
can be written as follows
\beq
\cZ=\cF^\La \alpha_\La-\cG_\La \beta^\La
\label{qtre}
\eeq
with the symplectic (real) vector field strength as defined by eq.
(\ref{tset}). It is easy to show that
\beq
\int_{S_2\times B}Z=n^\La_m\ \ ,\ \ \int_{S_2\times A} Z=n^e_\La
\label{qquabis}
\eeq
and, denoting by $*$ the Hodge dual,
\beq
\cZ^*=\cZ
\label{qqua}
\eeq
due to the property\cite{suzuki}
\begin{eqnarray}
\alpha^* &=& A\alpha +B\beta\ \ \ \alpha^{**}=-\alpha\nonumber\\
\beta^* &=& C\alpha +D\beta\ \ \ \beta^{**}=-\beta\ ,
\label{qcin}
\end{eqnarray}
where
\beq
\cS=\pmatrix{D&C\cr B&A\cr}
\label{qsei}
\eeq
is a symplectic matrix with entries
\begin{eqnarray}
A &=&-D^T=\re\cN (\im\cN)^{-1}\nonumber\\
C &=&(\im\cN)^{-1}\nonumber\\
B &=& -\im\cN-\re\cN(\im\cN)^{-1}\re\cN
\label{qset}
\end{eqnarray}
and satisfying $\cS^2=-\bfone$. From eqs. (\ref{qdue}) and (\ref{qcin})-(
\ref{qset})
it also follows that
\beq
\Omega^*=i\bar\Omega\ \ ,\ \ D_i\Omega^*=-i\, D_{\bar \imath}\bar\Omega\ .
\label{qott}
\eeq
It is easy to see that the matrix $\cS$ is related to the symplectic matrix
$\cM$ defined by eq. (\ref{tven}) by the relation
\beq
\cM=\cI\cS\ \ ,\ \ \cI=\pmatrix{0&-\bfone\cr\bfone&0\cr}\ .
\label{qnov}
\eeq
The real cohomology basis in $H^3$ can be decomposed in the Dolbeault
cohomology basis
\beq
H^3=H^{(3,0)}+H^{(2,1)}+H^{(1,2)}+H^{(0,3)}\ .
\label{qdie}
\eeq
This corresponds to introducing
\beq
\Omega \ \ ,\ \ \bar\Omega\ \ ,\ \ D_i\Omega\ \ ,\ \ D_{\bar \imath} \bar\Omega
\label{qund}
\eeq
with the following  vanishing intersections
\beq
\int\Omega\wedge D_i\Omega=\int\Omega\wedge D_{\bar \imath}\bar \Omega=0\ ,
\label{qdod}
\eeq
where
$D_i\Omega=\del_i\Omega-{{<\del_i \Omega,\bar\Omega>}\over
{<\Omega ,\bar\Omega>}}$, $i<\Omega,\bar\Omega>=e^{-K}$, $G_{i\jb}=-i
<D_i \Omega, D_\jb \bar\Omega> e^K$. It is easy to see that
$e^{K/2} \Omega\to V$, $e^{K/2} D_i \Omega\to U_i$ as defined in section 2
by eqs. (\ref{ddue}), (\ref{dtre}).
If one defines the auxiliary five-form
\begin{eqnarray}
&&\cZ^-=\cF^{-\La}\alpha_\La-\cG^-_\La\beta^\La=\nonumber\\
&&i\,e^{K/2} (T^-\bar\Omega+\cF^{-i} D_i\Omega)
\label{qtred}
\end{eqnarray}
(the above identity follows from (\ref{tott})-(\ref{tdie}) ), then
\cite{BCDFF,FBC}
\begin{eqnarray}
\cZ&=&{i\over2}e^{K/2}(T^-\bar\Omega-T^+\Omega\nonumber\\
&+&
\cF^{-i} D_i\Omega -\cF^{+\bar \imath} D_{\bar \imath} \bar \Omega )
=\re \cZ^-\ .
\label{qquatt}
\end{eqnarray}
We then see that from eq. (\ref{qquatt}) and (\ref{tund}),
 (\ref{tdod}) it follows
that
\begin{eqnarray}
&&\int_{S_2\times C} Z\wedge \Omega=-\half e^{-K/2}\int_{S_2} T^-\nonumber\\
&&=X^\La n_\La^e-F_\La n^\La_m=Z^{(3,0)}\ ,
\label{qquin}\\
&&\int_{S_2\times C} Z\wedge D_i \Omega=-\half e^{-K/2} \int_{S_2}\cF^{+\jb}
G_{i\jb}\nonumber\\
&&=D_i X^\La n^e_\La-D_i F_\La n^\La_m=Z_i^{(2,1)}\ .
\label{qsed}
\end{eqnarray}
The charges defined by eqs. (\ref{qquin}), (\ref{qsed})
$(Z^{(3,0)},Z_i^{(2,1)})$
 define the
physical charges with respect to the vector field strengths which are
 supermultiplet eigenstates. In fact, $(Z^{(3,0)},
Z^{(0,3)},Z^{(2,1)}_i,Z^{(1,2)}_{\bar\imath})$ correspond
to electric and magnetic charges of the $U(1)^{h_{21}+1}$ gauge group of the
complex cohomology. We also note that the 4D antiselfdual part of the five-form
$\cZ$ decomposes under $H^{(0,3)}+H^{(2,1)}$, due to the helicity
properties of the gravitino and gaugino.

Note that in the Type II string theories these charges are nonperturbative in
nature\cite{huto1,stromco}
 since all the gauge fields are Ramond-Ramond vectors (coming from
the five-form). The charges $(Z^{(3,0)},Z_i^{(2,1)})$
are then due to the solitonic
three-brane configurations. The BPS states have masses given by
\beq
M=e^{K/2} \mid Z^{(3,0)}\mid =\mid Z\mid\ .
\label{qdic}
\eeq
The charges $(Z^{(3,0)}, Z_i^{(2,1)})$ satisfy sum rules given by eqs.
 (\ref{tdic}), (\ref{tdoc}).
These charges  are the generalization to
Calabi-Yau manifolds of particular
combinations $Q^2_R$ and $Q^2_L$ of string theories. Black hole level
matching\cite{FHSV}
 should be compared to a quantization condition between these quantities,
as shown in the examples at the end of section 3.

Many of the topics covered in this report were obtained in collaboration with
M. Bill\`o, P. Fr\`e, T. Regge, P. Soriani and A. Van Proeyen whom we would
like to thank.

\end{document}